\documentstyle[seceq,preprint,epsf]{ptptex}
\newcommand{\nn}{\nonumber}
\newcommand{\Slash}[1]{\ooalign{\hfil/\hfil\crcr$#1$}}
\newcommand{\QED}{\mbox{QED}_3}
\newcommand{\psibar}{\bar{\psi}}
\newcommand{\chibar}{\bar{\chi}}
\newcommand{\Ncr}{N_{\mbox{\small cr}}}
\newcommand{\Os}{{\cal O}_{\mbox{\tiny S}}}
\newcommand{\Op}{{\cal O}_{\mbox{\tiny P}}}
\newcommand{\Gs}{G_{\mbox{\tiny S}}}
\newcommand{\Gp}{G_{\mbox{\tiny P}}}

\preprintnumber[4cm]{%<-- [..]: optional width of preprint # column.
KANAZAWA-00-13\\December, 2000}
\title{%        %You can use \\ for explicit line-break
Dynamical Symmetry Breaking in $\QED$ from the Wilson RG Point of View
}
\author{%       %Use \sc for the family name
Ken-Ichi {\sc Kubota}
\footnote{E-mail address: kubota@hep.s.kanazawa-u.ac.jp} 
and 
Haruhiko {\sc Terao}
\footnote{E-mail address: terao@hep.s.kanazawa-u.ac.jp}
}
\inst{%         %Affiliation, neglected when [addenda] or [errata]
Institute for Theoretical Physics, Kanazawa University,\\
Kanazawa 920--1192, Japan
}
\recdate{%      %Editorial Office will fill in this.
}
\abst{%       %this abstract is neglected when [addenda] or [errata]
Dynamical symmetry breaking in three dimensional QED with $N$
flavors, 
which has been mostly analyzed by solving the Schwinger-Dyson
equations,
is investigated by means of the approximated Wilson, or
non-perturbative, renormalization group (RG).
We study the RG flows of the gauge coupling and the general four-fermi
couplings allowed by the symmetry with concentrating our interest
on study of the phase structure.
The RG equations have no gauge parameter dependence in our
approximation scheme.
It is found that there exist chirally broken and unbroken phases
for $N > N_{\rm cr}$ ($3 < N_{\rm cr} < 4$) and that the unbroken phase
disappears for $N < N_{\rm cr}$.
We also discuss the spontaneous parity breaking in $\QED$ with the
four-fermi interactions.
}

\begin{document}

\maketitle

\section{Introduction}
Dynamical symmetry breaking in (2+1)-dimensional massless 
Quantum Electrodynamics ($\QED$) has been attracting much attention 
since Pisarski and  Appelquist {\it et al.}\cite{first,ladder}
have found the novel chiral symmetry breaking depending on the 
number of flavors by solving the approximated Schwinger-Dyson 
(SD) equations.
In the case that $\QED$ contains $N$ flavors of four-component spinors,
or $2N$ flavors of two-component spinors, 
the global symmetry is enhanced to $U(2N)$ rather than $U(N)$.
This enhanced symmetry may be regarded as a sort of chiral symmetry 
even though the theory is defined in three-dimensions.
Indeed this symmetry is spontaneously broken by dynamical mass
generation.

In these analyses the photon self-energy evaluated in the large $N$ limit is
inserted into the gap equation obtained in the so-called ladder
approximation.
It has been claimed in this way \cite{ladder} that 
the chiral symmetry is spontaneously
broken for $N < \Ncr = 32/\pi^2 \sim 3.24$, while unbroken for $N > \Ncr$.
In practice the SD equations derived in the ladder approximation 
suffer from large gauge parameter dependence.
The above result is brought by using the Landau gauge.

Subsequently a lot of works have been devoted for
improvement of the approximation schemes for the SD equations and
for studies of various aspects of the phase transition \cite{nash,others,order}.
In ref.~\citen{nash}, the approximation is refined so that the gap equations
is made gauge parameter independent by proceeding to the next-to-leading
order of $1/N$ expansion.
Later the effects of the wave function renormalization ignored in the
original approximation have been also incorporated.
Then the coupled SD equations for the photon self-energy and the mass 
function were examined.
As results, such elaborate treatment of the SD equations also
supports the qualitative picture of the chiral symmetry breaking
mentioned above. Even the critical flavor number has not been altered
significantly.

On the other hand  numerical simulations have been also performed
for non-compact $\QED$ defined on the lattice \cite{lattice}.
It seems remarkable that the results from these simulations 
are consistent with those obtained by solving the approximated
SD equations, namely $3<\Ncr<4$.

$\QED$ is invariant also under parity transformation as long as
the Chern-Simons term is absent.
It has been shown by Redlich \cite{Redlich} that the Chern-Simons
term should be generated through gauge invariant regularization
with odd number of two-component massless spinors.
However it is expected according to Vafa-Witten's argument \cite{VafaWitten}
that the parity symmetry is not spontaneously broken for $\QED$ with four-component
spinors.
Actually it was seen \cite{parity} 
that the analyses using the SD equations are
consistent with this expectation by considering the gap equations for
the chirally invariant but parity odd dynamical mass.

After that these studies were extended to $\QED$ with fermion self-interactions
\cite{4fermiQED}.
This was motivated by the expectation that the parity symmetry can be
spontaneously broken 
in the presence of general four-fermi interactions.
The SD analyses show the presence of parity broken phase and also suggest
that this phase is separated from the chiral symmetry broken phase.
The Chern-Simons term is generated through radiative correction in the
parity broken phase even in the case of even number of flavors.

Apart from this, the dynamics of $\QED$ with the Chern-Simons term has been also 
intensively discussed \cite{CS,Hosotani},
specially in view of its applications in modelling (high-temperature)
superconductors \cite{superconductor,finitetemp}.
In this paper, however, we consider only $\QED$ with even number of flavors and
do not include the Chern-Simons term.
Furthermore the three dimensional Thirring model
were also examined by numerical simulations
as well as by solving the SD equations \cite{Thirring}.

The Exact Renormalization Group (ERG) \cite{Wilson,WH,Polchinski,Legendre,ERG}, 
which represents continuous version of the Wilson RG transformation, 
has been known as an analytical method applicable to non-perturabative 
dynamics of field theories.
Both of the ERG and the SD equations are given in functional forms
and lead to the correlation functions as their solutions.
It is also common that regularization is necessary and that
some approximations are inevitable in practical calculations,
but it should be noted that the ERG equations give the RG flows for the 
effective couplings, while the SD gives order parameters in terms of 
bare parameters.
\footnote{The relations between ERG and SD equations are discussed in 
ref.~\citen{terao}.}
In analysis shown in this paper, we will perform rather brute
approximations. Therefore let us call the approximated ERG as
Non-perturbative (NP) RG hereafter.

NPRG has not been applied to clarify the phase structures of
dynamical symmetry breaking in $\QED$ (with the four-fermi interactions)
in spite of much interest mentioned above.
\footnote{In ref.~\citen{PRG} 
dynamical symmetry breaking of four-fermi theories
are examined by applying the RG method to their effective (composite) theories.
}
The purpose of this paper is to show that we may clarify the phase structures
of these theories much easier than the SD approaches.
It will be clearly seen by NPRG that presence of the IR attractive fixed 
point is essential for the novel phase transition.
Moreover we can directly find out the boundaries between the chiral
symmetry broken phase, the parity broken phase and the unbroken phase
just by following the RG flows.
This point is a great contrast to the SD approach.
Note that in the SD approach we need to assume the order parameters for
the symmetries to be broken apriori to derive the gap equations and to solve
them for every theory.
In the RG approach, however, we may treat any theories invariant under
the symmetries on the equal footing without concerning the order
parameters.
We aim to demonstrate such advantageous points of NPRG
in comparison with the SD analyses through explicit calculations 
for $\QED$, not to pursue further improvement of the 
approximations.

Dynamical chiral symmetry breaking in (3+1)-dimensional gauge theories
has been analyzed by NPRG in the case of a single massless flavor 
\cite{Aoki1,Aoki2,Sumi}.
Four-fermi interactions are induced in the effective theories by the
gauge interactions.
It was found that the RG flows of the four-fermi couplings reveal the phase 
structure.
Also if we truncate the corrections for the four-fermi interactions 
up to a set called the ladder-type in the later discussions, 
then the phase boundary obtained by solving the SD equation in the ladder
approximation is found to be precisely reproduced.
The anomalous dimensions of fermion composite operators are immediately
calculated from the RG equations.

Moreover the gauge parameter dependence is remarkably improved by
using the NPRG equations obtained by the derivative expansion \cite{Aoki1}.
The momentum cutoff introduced to define the Wilson RG breaks
the gauge invariance. The remained gauge dependence comes purely from this 
cutoff procedure, while the ladder approximation badly destroys the gauge 
invariance. We may remove the gauge dependence by substituting the 
anomalous dimension of the fermion with one evaluated in a gauge invariant way.
We will adopt such an approximation scheme in the RG analyses of $\QED$
as well.

The rest of our paper is organized as follows.
In section~2, we briefly review the NPRG for a simple four-fermi
theory and how dynamical symmetry breaking is described in the RG
framework.
Section~3 is our main part. 
First we consider the general four-fermi operators allowed
by the symmetries of $\QED$.
After explaining our approximation scheme to the NPRG,
we explicitly examine the phase structure by looking at the
RG flows with varying the number of flavors.
In the Wilson RG framework we treat all the effective theories
at the same time.
In our scheme the phases of $\QED$ with general four-fermi
interactions are explored automatically.
Thus we can discuss not only the chiral symmetry breaking but also
the parity breaking.
Lastly the broken symmetries are discussed in section~4.
Actually the RG flows themselves tell about the phases but not
the symmetries.
We discuss a simple way to know the broken symmetry by utilizing the
RG flows.
Secton~5 is devoted to conclusions.

\section{RG equation for the four-fermi coupling and chiral symmetry breaking }
Here we consider the large $N$ Gross-Neveu model as the simplest example for
the dynamical chiral symmetry breaking \cite{Aoki1,Aoki2}.
Of course this model can be analyzed easily by considering the gap equation.
Let us see how we can obtain the same results by solving the NPRG 
equations.

Wilson RG observes variation of the effective actions in lowering the
cutoff scale.
Since any operators allowed by the original symmetries are
generated in the effective action through the radiative corrections,
we need to consider infinitely many effective couplings.
In the large $N$ limit, however, the ERG equation for the cutoff effective
potential $V(\psi, \bar{\psi}; \Lambda)$ 
may be exactly derived.
The approximation truncating any momentum dependent corrections
is called the Local Potential Approximation (LPA) \cite{LPA}.
Namely the LPA becomes exact in the large $N$ limit.

In the later discussion we want to compare the results with
those obtained by the SD equations defined with sharp cutoff.
Therefore let us adopt the sharp cutoff also for the RG equations.
It has been also known \cite{sharp} that the sharp cutoff limit 
of ERG equations reduce to the Wegner-Houghton RG equation \cite{WH}.
In this formulation cutoff is performed to the path integration measure:
\begin{equation}
Z =  \int \prod_{|p| < \Lambda} 
{\cal D}\psi(p){\cal D}\bar{\psi}(p) 
e^{-S_{\mbox{\scriptsize eff}}[\psi, \bar{\psi}; \Lambda]}.
\end{equation}
Here we may set the eucledian effective action for the large 
$N$ Gross-Neveu model in $D$ dimensions as
\begin{equation}
S_{\mbox{\scriptsize eff}} = 
\int d^D x~
\bar{\psi}_i \gamma_{\mu}\partial_{\mu}\psi_i
+V(\sigma),
\end{equation}
where $\sigma$ denotes a product of the classical fields, 
$\bar{\psi}\psi$.
It should not be confused with the expectation value of
the fermion composite, $\langle \bar{\psi}\psi \rangle$.

The ERG for the effective potential is given with the
scale parameter $t=\ln(\Lambda_0/\Lambda)$ by
\begin{equation}
\frac{d V(\sigma)}{dt}=
DV(\sigma) - (D-1)V'(\sigma)- A_D \ln\left(1 + V'(\sigma)^2\right),
\end{equation}
where the prime stands for the derivative with respect to $\sigma$ and $A_D$
is a constant depending on the space-time dimensions.
It is found \cite{Aoki2} that the dynamical mass treated by the SD equations
is obtained from this effective potential as
\begin{equation}
m_{\rm eff} 
= \lim_{\Lambda \rightarrow 0} V'(\sigma,\Lambda) |_{\sigma=0}.
\end{equation}

The ERG has a great advantage to find the phase structures
and also the critical exponents compared with the SD approaches. 
If we perform the operator expansion of the effective potential
into
\begin{equation}
V(\sigma; \Lambda)=-\frac{1}{2\Lambda^{D-2}} G(\Lambda)\sigma^2
+ \frac{1}{8\Lambda^{3D-4}}G_8(\Lambda) \sigma^4 + \cdots,
\end{equation}
we may derive beta functions for each coupling.
Then the effective four-fermi coupling $G$ is found to be 
subject to the ERG equation isolated from others:
\begin{equation}
\beta_{G}=\Lambda\frac{dG}{d\Lambda}
= (D-2)G - A_D G^2.
\end{equation}
This beta function has two fixed points:
$G^*=0$ (IR attractive) and $G^*=(d-2)/A_D$ (IR repulsive).
The IR repulsive fixed point gives the critical coupling
of the chiral symmetry breaking.
Thus we can immediately find from the RG flows that there exist
two phases; broken and unbroken ones. 
It is also quite easy
to calculate the anomalous dimensions of the operators
$\bar{\psi}\psi$, $(\bar{\psi}\psi)^2$ and so on by this method.

It would be important to note that the mass term or
any symmetry breaking operator does not
appear in the effective action even in the chirally
broken phase since the RG evolution respects the original
symmetries.
However if we solve the ERG equation for the effective potential
$V(\sigma;\Lambda)$, then it is found that
the potential is evolved to be non-analytic at the origin
due to IR singularity of massless fermion loops \cite{Aoki2}.
Thus the dynamical mass generation is observed in a rather
non-trivial way.
As is shown in ref.~\citen{Aoki2}, it is practically useful to 
introduce collective coordinates corresponding to the
fermion composites into the effective action in order to
evaluate the order parameters.
However it should be noted that 
the RG flows of the four-fermi couplings are unable to
conclude the broken symmetries even though the phase boundaries
are exposed by them.
In section~4 we are coming back to this problem in order to see
the symmetry spontaneously broken in $\QED$.

\section{NPRG for $\QED$ and dynamical symmetry breaking}

\subsection{Four-fermi interactions}

Let us consider QED$_3$ with $N$ flavors of four-component 
spinors, $\psi^i$ $(i=1, \cdots ,N)$.
The bare lagrangian is given in eucledian space by
\begin{equation}
{\cal L}_b
 = \frac14 F_{\mu \nu}^2 + 
   \psibar_i \gamma_{\mu} \left( \partial_{\mu} - i e A_{\mu} \right) \psi^i
   - \frac1{2 \xi} ( \partial_{\mu} A_{\mu})^2,
\label{psiversion}
\end{equation}
where we suppose that the Chern-Simons term is absent.
We use the 4 by 4 $\gamma$ matrices given by 
\begin{equation}
\gamma_0 = 
\left( 
\begin{array}{cc}
\sigma_3 & 0 \\
0 & -\sigma_3
\end{array}
\right),
\gamma_1 = 
\left( 
\begin{array}{cc}
\sigma_1 & 0 \\
0 & -\sigma_1
\end{array}
\right),
\gamma_2 = 
\left( 
\begin{array}{cc}
\sigma_2 & 0 \\
0 & -\sigma_2
\end{array}
\right),
\gamma_3 = 
\left( 
\begin{array}{cc}
0 & -i \\
i & 0 
\end{array}
\right).
\label{rmatrix1}
\end{equation}
Also we introduce 
$\gamma_5 =\gamma_0  \gamma_1  \gamma_2  \gamma_3$
and
$\tau= -i \gamma_5  \gamma_3$.

This lagrangian is invariant under the global $U(2N)$ and also the parity 
symmetry.
The parity transformation is defined by
$\psi^i(x) \mapsto \psi'^i(x')=i\gamma_3 \gamma_1 \psi^i(x)$ for $x'=(t,-x,y)$.
$U(2N)$ symmetry is made more transparent by reformulating in terms of $2N$
two-component spinors, $\chi^I$ $(I=1,\cdots, 2N)$:
\begin{equation}
\psi^i = 
\left(
\begin{array}{c}
\chi^i \\
\chi^{i+N}
\end{array}
\right),~~~~
\bar{\psi}_i = 
(\chi_i^{\dagger}\sigma_3, -\chi_{i+N}^{\dagger}\sigma_3)
=(\bar{\chi}_i, \bar{\chi}_{i+N})\tau.
\end{equation}
The two-component fields are transformed by the $U(2N)$ matrix $U$ as
$\chi^I \mapsto \chi'^I= U_J^I \chi^J $.
Therefore 
$\bar{\psi}_i \gamma_{\mu} \psi^i =
\bar{\chi}_I \sigma_{\mu}\chi^I$
is invariant under the both symmetries.
The ordinary mass operator,
$\bar{\psi}_i \psi^i = 
\bar{\chi}_i \chi^i - \bar{\chi}_{i+N}\chi^{i+N}$, 
is parity even but not invariant under the $U(2N)$ transformation.
If this operataor acquires a non-vanishing vacuum expectation value,
then $U(2N)$ is spontaneously broken to $U(N) \times U(N)$.
Thus we may regards this $U(2N)$ symmetry as a sort of
chiral transformation.
While we find a $U(2N)$ invariant operator,
$\bar{\psi}_i \tau \psi^i = 
 \bar{\chi}_I \chi^I $,
it is parity odd in turn.
Therefore non-vanishing expectation value of this operator
leads to spontaneous breakdown of the parity symmetry.
However it is expected from the Vafa-Witten theorem
that the parity symmetry is never broken in QED$_3$.

In section~2 we saw that the RG flows of the effective four-fermi 
interactions are important to distinguish the phases.
We can list up all the local four-fermi operators invariant under 
$U(2N)$ and parity transformations as follows:
\begin{eqnarray}
\Op&=&
(\bar{\psi}_i\tau \psi^i)^2 = (\bar{\chi}_I\chi^I)^2 \\
{\cal O}_{\mbox{\tiny V}} &=&
(\bar{\psi}_i\gamma_{\mu} \psi^i)^2 
= (\bar{\chi}_I\sigma_{\mu}\chi^I)^2 \\
\Os &=& \frac12 \left[
\bar{\psi}_i \psi^j\bar{\psi}_j \psi^i 
- \bar{\psi}_i \gamma_3 \psi^j \bar{\psi}_j \gamma_3\psi^i  
- \bar{\psi}_i \gamma_5 \psi^j \bar{\psi}_j \gamma_5\psi^i 
+ \bar{\psi}_i \tau \psi^j \bar{\psi}_j \tau \psi^i \right]\nn \\
 &=& \bar{\chi}_I\chi^J \bar{\chi}_J\chi^I  \\
{\cal O}_{\mbox{\tiny V}{}'} &=&
\bar{\psi}_i \gamma_{\mu}\psi^j\bar{\psi}_j \gamma_{\mu}\psi^i 
- 
\bar{\psi}_i \gamma_3\gamma_{\mu} \psi^j 
\bar{\psi}_j \gamma_3\gamma_{\mu}\psi^i \nn \\
& &- 
\bar{\psi}_i \gamma_5\gamma_{\mu} \psi^j 
\bar{\psi}_j \gamma_5\gamma_{\mu}\psi^i 
+ 
\bar{\psi}_i \tau \gamma_{\mu}\psi^j 
\bar{\psi}_j \tau \gamma_{\mu}\psi^i \nn \\
 &=& 2 \bar{\chi}_I\sigma_{\mu}\chi^J \bar{\chi}_J\sigma_{\mu}\chi^I 
\end{eqnarray}
These operators are induced by radiative corrections. However
it is found by the Fierz transformation 
that only two of them are independent.
We choose $\Os$ and $\Op$ as the independent ones
and always rewrite others by using the Fierz transformation
whenever they are induced.

\subsection{NPRG and the approximation scheme}
In this subsection we explain the outline of our approximation scheme.
First we truncate the set of induced operators in the effective lagrangian to
%effective lagrangian (chi version)
\begin{equation}
{\cal L}_{\mbox{{\scriptsize eff}}}
 = \frac14 F_{\mu \nu}^2 + 
   \chibar_I \left( \Slash{\partial} - i e \Slash{A} \right) \chi^I 
   - \frac1{2 \xi} ( \partial_{\mu} A_{\mu})^2
   - \frac{\Gs}2 \Os - \frac{\Gp}2 \Op.
\label{efflagrangian}
\end{equation}
%where $\Slash{\paritial}$ and $\Slash{\partial}$ 
Therefore the RG flows are given in the three dimensional coupling space
spanned by $(e^2, \Gs, \Gp)$.
As is seen in the previous section the RG equations of the four-fermi 
couplings are separated from other multi-fermi couplings in the LPA.
Namely we may obtain the same RG equations for the four-fermi couplings
if we perform the operator expansion to the effective potential
evaluated in the LPA.
The reason to truncate other operators is that the four-fermi couplings
are enough to explore the phase structures.
In the RG approach we may naturally incorporate all the theories
with the identical symmetries.
For example, we can consider
the model with a bare ${\cal O}_{\mbox{\tiny V}}$ operator, namely the
massless Thirring model, at the same time.

Since our purpose is  to see the chiral phase structure of
QED$_3$, we adopt the chiral $U(2N)$ symmetry and parity preserving 
regularization, {\it i.e.} naive momentum cutoff, at the
cost of the gauge invariance. 
Here we discard the gauge non-invariant corrections 
induced in such a regularization scheme, {\it e.g.} photon mass,
and substitute the beta function for the gauge coupling by
the one-loop perturbative one as the first step of 
approximation.
Of course the gauge invariant scheme \cite{gauge} is preferable to see
non-perturbative dynamics by gauge interactions in general. 
However, the manifest chiral symmetry would be necessarily lost
and we would face up to the problem to extract 
chirally invariant theories.
This is  similar to the problem appearing
in the lattice gauge theories. 
In this paper we do not pursue for this direction.

The Chern-Simons term
\begin{equation}
{\cal L}_{\rm CS}= \frac{i}{2} \theta \epsilon_{\mu \nu \rho}
A_{\mu} \partial_{\nu} A_{\rho} 
\end{equation}
cannot be generated in the Wilsonian effective action since such a correction
is forbidden by the parity symmetry.
On the other hand, however, in the parity broken phase
(which exists for the models with the bare four-fermi interactions), 
the fermions are supposed to acquire the parity breaking effective mass.
Therefore the CS term is expected to be dynamically generated 
through radiative corrections.
This contradicting situation is related to the appearance of
the dynamical mass in the RG framework.
It is expected that the Chern-Simons term will be generated 
once we revaluate the corrections at the parity broken vacua.
Here we leave this problem to the future investigations.

Now we consider the RG equations for the three couplings,
$(e^2, \Gs, \Gp)$.
It is not necessary to derive the NPRG for the effective action
to find them.
Indeed the corrections appearing in the ERG formulation are
given also by evaluating one-loop diagrams with internal
momentum scale fixed to $\Lambda$.

First the beta function for the gauge coupling may be obtained
in the above approximation scheme as
\begin{equation}
\frac{d e^2}{dt} = e^2 - \frac{N}{8}e^4,
\label{gbeta}
\end{equation}
where $t=\ln(\Lambda_0/\Lambda)$. 
The first term represents
the canonical scaling of the gauge coupling with dimension
one half.

In the original SD analyses \cite{first,ladder} the photon
self-energy was evaluated in the large $N$ leading order as
$\Pi(p^2) = N e_0^2/8 \sqrt{p^2}$ with
the bare gauge coupling $e_0$.
Note that the self-energy is finite in three dimensions.
The photon propagator in the gap equation is defined with
this self-energy.

We may define also the renormalized gauge coupling as
\begin{equation}
e^2(p) = \frac{e^2_0}{1+\Pi(p^2)}.
\label{relation}
\end{equation}
If we identify the renormalization scale $\mu$ with the momentum $p^2$,
then this renormalized coupling satisfies the above RG equation.
It should be noted that there
appears an IR stable fixed point of the gauge coupling in three dimensions.
As is seen later this fixed point plays an essential
role for the novel phase transition in the RG point of view.
The fixed point coupling is given by $e^2{}^* = 8/N$.
When $N$ is not large, this appears at the strong coupling
region. 
Therefore the present perturbative treatment is not justified actually.
However, as long as the fixed point structure of the gauge beta function is
not altered significantly, the essential mechanism of the dynamical
symmetry breaking is thought to be captured in this simple
RG treatment. 
The evaluation of the gauge beta function reliable in the strong
coupling region remains as an open problem.

The beta functions for the four-fermi couplings $\Gs$ and $\Gp$
are evaluated by summing up the corrections described in Fig.~1.
\begin{figure}[h]
\begin{center}
\epsfxsize=0.7\textwidth
\leavevmode
\epsffile{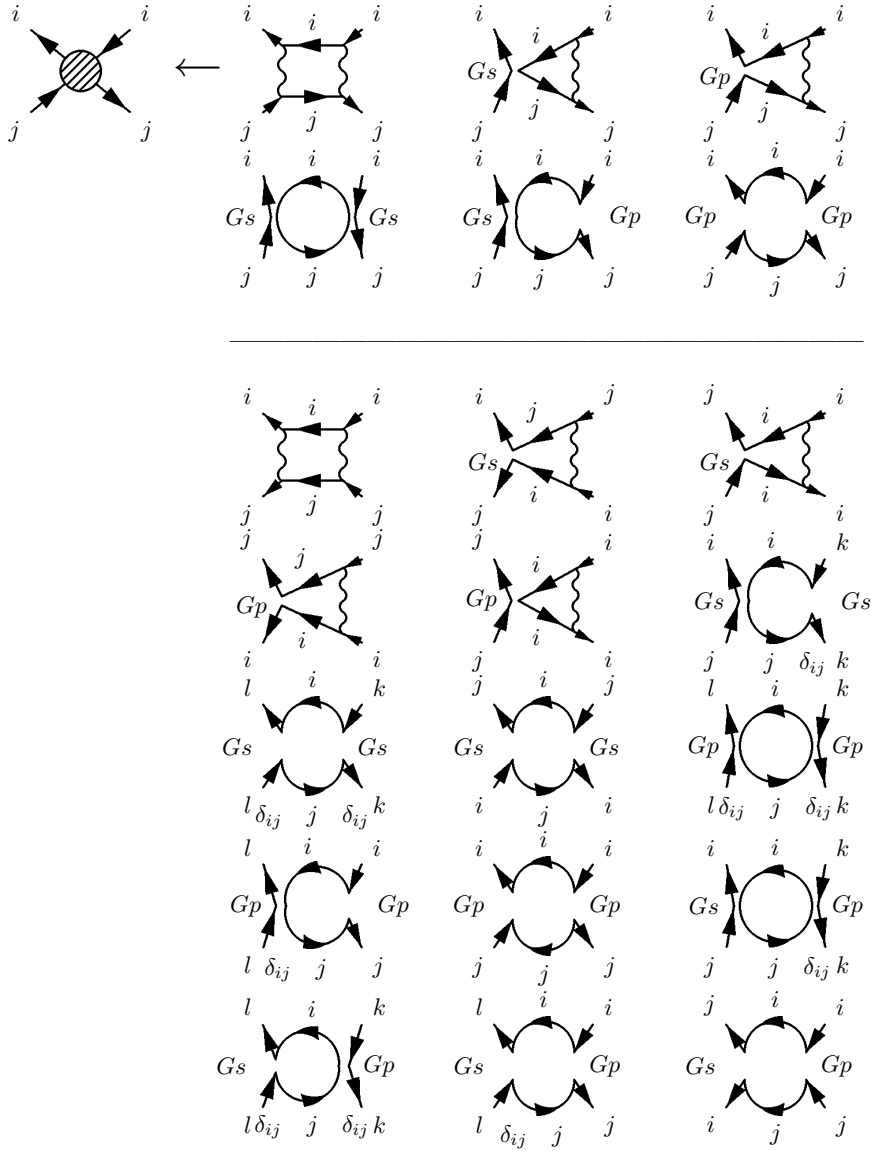}
\caption{ The Feynman diagrams for the corrections to the four-fermi operators are shown.
The arrows represent contraction of the spinor indices.}
\end{center}
\end{figure}
At the vertices the operators corresponding to the couplings are
inserted.
We call the diagrams in the first two lines of Fig.~1 ``ladder type''
and the others ``non-ladder type''.
When we restrict ourselves ``ladder type diagram'', the results obtained 
in the ladder SD equation are precisely reproduced as we will see below.
We examine the RG flows explicitly and discuss the phase structures
in the case including only the ladder type corrections and
the full corrections separately  in the next two subsections.

\subsection{The ladder approximation}
We also adopt the Landau gauge propagator to evaluate the
ladder type diagrams.
Then the beta functions for the four-fermi couplings $\Gs$
and $\Gp$ are easily found out to be
\begin{eqnarray}
\dot{\Gs} &=& -\Gs + \frac1{\pi^2}
               \left[
	        \Gs{}^2 - \Gs \Gp + \frac13 \Gp{}^2 +
	        2 e^2 \Gs -\frac43 e^2 \Gp +\frac23 e^4 
	       \right], 
\label{ladderRGE1}\\
\dot{\Gp} &=& -\Gp + \frac1{\pi^2}
               \left[
	        \frac16 \Gp{}^2 -
	        \frac23 e^2 \Gp - \frac23 e^4 
	       \right],
\label{ladderRGE2}
\end{eqnarray}
where dot in the left hand side stands for the derivative with
respect to $t = \ln (\Lambda_0/\Lambda)$.
By defining a new variable, $\Gs' = \Gs - \Gp/2$, 
we can separate the beta function for $\Gs'$ from that for $\Gp$
as
\begin{equation}
\dot{\Gs'} = -\Gs' + \frac1{\pi^2}
              \left[
	       \Gs'{}^2 + 2 e^2 \Gs'+ e^4
	      \right].
\label{ladderRGE}
\end{equation}

Now we may solve the coupled equations Eq.~(\ref{gbeta}) and Eq.~(\ref{ladderRGE}).
It is easily found that there exist two fixed points, $\Gs'{}^{(\pm)}$, 
\begin{equation}
 \Gs'{}^{(\pm)} = \frac12\left(1-\frac{2e^2{}^*}{\pi^2}\right)
 \pm \sqrt{\frac14 - \frac{e^2{}^*}{\pi^2}},
\label{ladderFP}
\end{equation}
only if  $e^2{}^* < \pi^2/4$, {\it i.e.} $N > \Ncr=32/\pi^2$.
At the critical number, these two points
merge each other and there is no fixed point solution
for $N < \Ncr$.

$\Gs'{}^{(-)}$ ($\Gs'{}^{(+)}$) is an IR (UV) fixed point respectively.
Existence of the IR fixed point, $\Gs'{}^{(-)}$, is important as far as 
the dynamical symmetry breaking in $\QED$ is concerned.
The RG flow diagrams in ($e^2, \Gs'$)-space 
are shown in Fig.~2 and Fig.~3 in the case of $N = 4$ and $2$
respectively.
Note that $\QED$ corresponds to flows starting from the $\Gs'=0$ line.
In Fig.~2 we see that the IR fixed point exists indeed.
In the asymptotically free region, all the flows are absorbed into the IR
fixed point, which means the theory is in the symmetric phase.
On the other hand, in the asymptotically non-free region, there appears
a phase boundary.
This indicates  that dynamical chiral symmetry breaking occurs 
even if $N > \Ncr$, provided the bare gauge coupling is strong enough.
As is seen in Fig.~3,
all the flows in both the regions blow up for $N < \Ncr$.
>From this behavior we may suppose that the chiral symmetry is
spontaneously broken irrespective of the bare gauge coupling.

\begin{figure}[h]
\begin{center}
\begin{minipage}[t]{60mm}
\epsfxsize=1.0\textwidth
\leavevmode
\epsffile{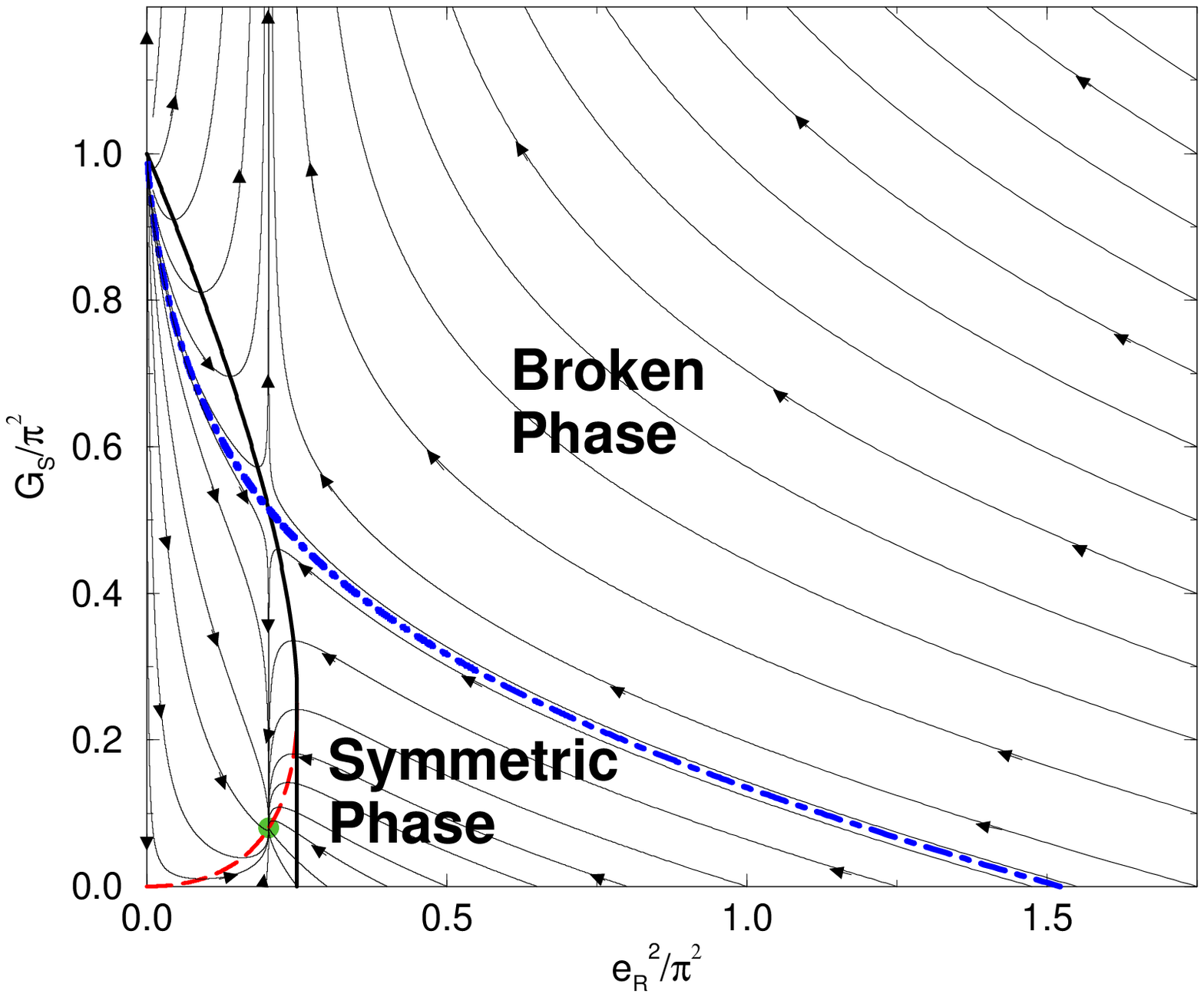}
\caption{RG flows for $\QED$ with $N=4$ flavors in the ladder approximation
are described on the $( e^2, \Gs )$-plane.
The dashed-dot line is the chiral phase boundary.}
\end{minipage}
\hspace{5mm}
\leavevmode
\begin{minipage}[t]{60mm}
\epsfxsize=1.0\textwidth
\epsffile{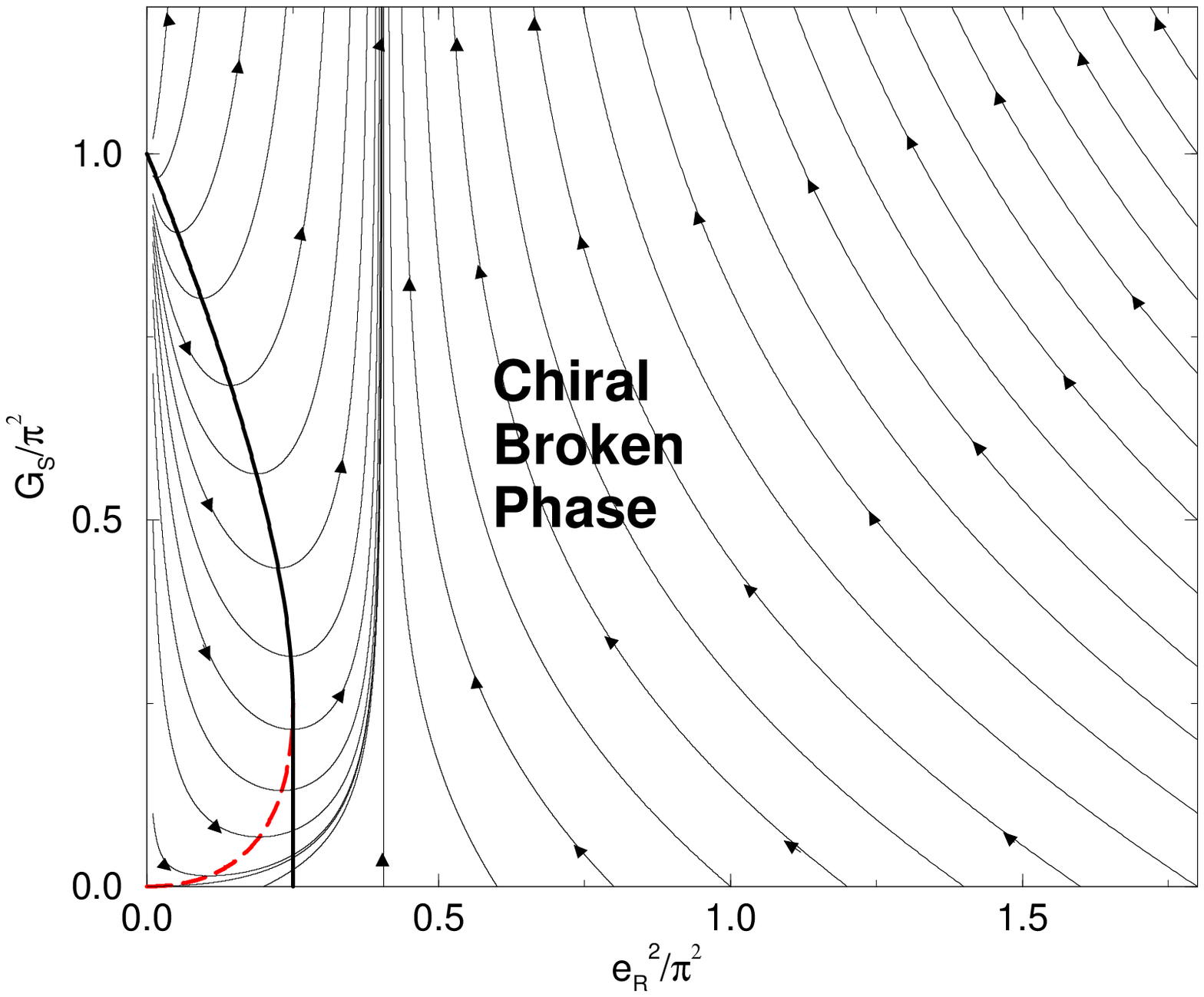}
\caption{RG flows for $\QED$ with $N=2$ flavors in the ladder approximation
are described on the $( e^2, \Gs )$-plane.}
\end{minipage}
\end{center}
\end{figure}

Our conclusion may sound slightly different from the results advocated
in ref.~\citen{first}, though the critical number of flavor just
coincides.
It was claimed that there is no non-trivial solution of the gap equation
as far as $N > \Ncr$ no matter how strong the gauge coupling is. 
The difference is thought to lie in the way of renormalization of the
gauge coupling.
If we make 
the renormalized coupling defined by Eq.~(\ref{relation})
dimensionless by scaling $e^2 \rightarrow e^2\mu$,
then $e^2$ never exceeds the fixed point value.
Namely the SD analysis examined only the asymptotically free region
of Fig.~2 and Fig.~3 in the RG point of view.
Therefore the SD result is not contradicting with our observation.

In the RG approach we put cutoff for any radiative corrections.
The effective gauge coupling obtained by solving Eq.~(\ref{gbeta})
is given by
\begin{equation}
e^2(\Lambda)= 
 \frac{e_0^2}{1 + \frac{N e^2_0}{8 \Lambda} -\frac{N e_0^2}{8 \Lambda_0}}.
\label{solution}
\end{equation}
This effective coupling can be made bigger than the IR fixed point,
$e^2{}^{*}$, 
owing to the bare cutoff scale.
On the other hand the renormalized coupling defined with the 
finite self energy corresponds to the infinite limit of the
bare cutoff.
Namely it may be said that the SD analysis restricts $\QED$ which
possesses continuum limit.
In the lattice simulations as well 
the continuum limit is observed \cite{lattice}.
Reversely we cannot take the continuum limit of the model
with the gauge coupling larger than $e^2{}^{*}$.
However we should be allowed to treat $\QED$ as an effective
theory with some UV cutoff.
Then there are found two phases.

\subsection{Results by the gauge independent RG equations}
Now let us investigate the RG flows taking account of the full 
corrections  to the four-fermi couplings given by Fig.~1. 
Both of the contributions from the ladder diagrams and from 
the non-ladder diagrams amount to the same order.
Therefore we should include the non-ladder ones as well.
However the most benefit to include them is recovering the
gauge independence lost in the ladder approximation.
The gauge parameter in the $e^4$ order contributions are found to 
cancel each other.
The gauge parameter dependence in the $e^2 \Gs (\Gp)$ order
should be eliminated by considering the fermion anomalous dimensions
$\eta$ \cite{Aoki1}.
Since the anomalous dimension cannot be evaluated in the LPA,
we must proceed to the derivative expansion.
Here, however, we simply substitute the perturbative result for 
$\eta$:
\begin{equation}
\eta= -\left(\frac{16-3N\xi}{6N\pi^2}\right)e^2.
\label{eta}
\end{equation}

Then the beta functions for $\Gs$ and $\Gp$ are found to be
\begin{eqnarray}
\dot{\Gs} &=& -\Gs + \frac1{\pi^2}
               \left[
	        \frac{N+2}{3} \Gs{}^2 - \Gs \Gp +
	        \frac43 e^2 \Gs -\frac83 e^2 \Gp 
	       \right],
\label{nonladderRGE1}\\
\dot{\Gp} &=& -\Gp + \frac1{\pi^2}
               \left[
	        (2 N-1) \Gp{}^2 - 2 (N-1) \Gs \Gp + \frac{4 N -7}{6} \Gs{}^2
	       \right. \nonumber \\
          & &  \left. \hspace{6cm}
	      -\frac83 e^2 \Gs + \frac43 e^2 \Gp - 2 e^4   
	       \right], 
\label{nonladderRGE2}
\end{eqnarray}
which are free from the gauge parameter.

In this case we cannot separate $\Gs$ part by the redefinition.
Therefore we examine these coupled differential equations including
Eq.~(\ref{gbeta}) numerically.
The RG flows run in the three dimensional theory space.
In Fig.~3 and Fig.~4 the RG flows for $\QED$, or starting from
$\Gs=\Gp=0$, are projected on the $(e^2, \Gs )$-plane.
Fig.~4 is for $N=4$ and Fig.~5 is for $N=2$. 
\begin{figure}[h]
\begin{center}
\begin{minipage}[t]{60mm}
\epsfxsize=1.0\textwidth
\leavevmode
\epsffile{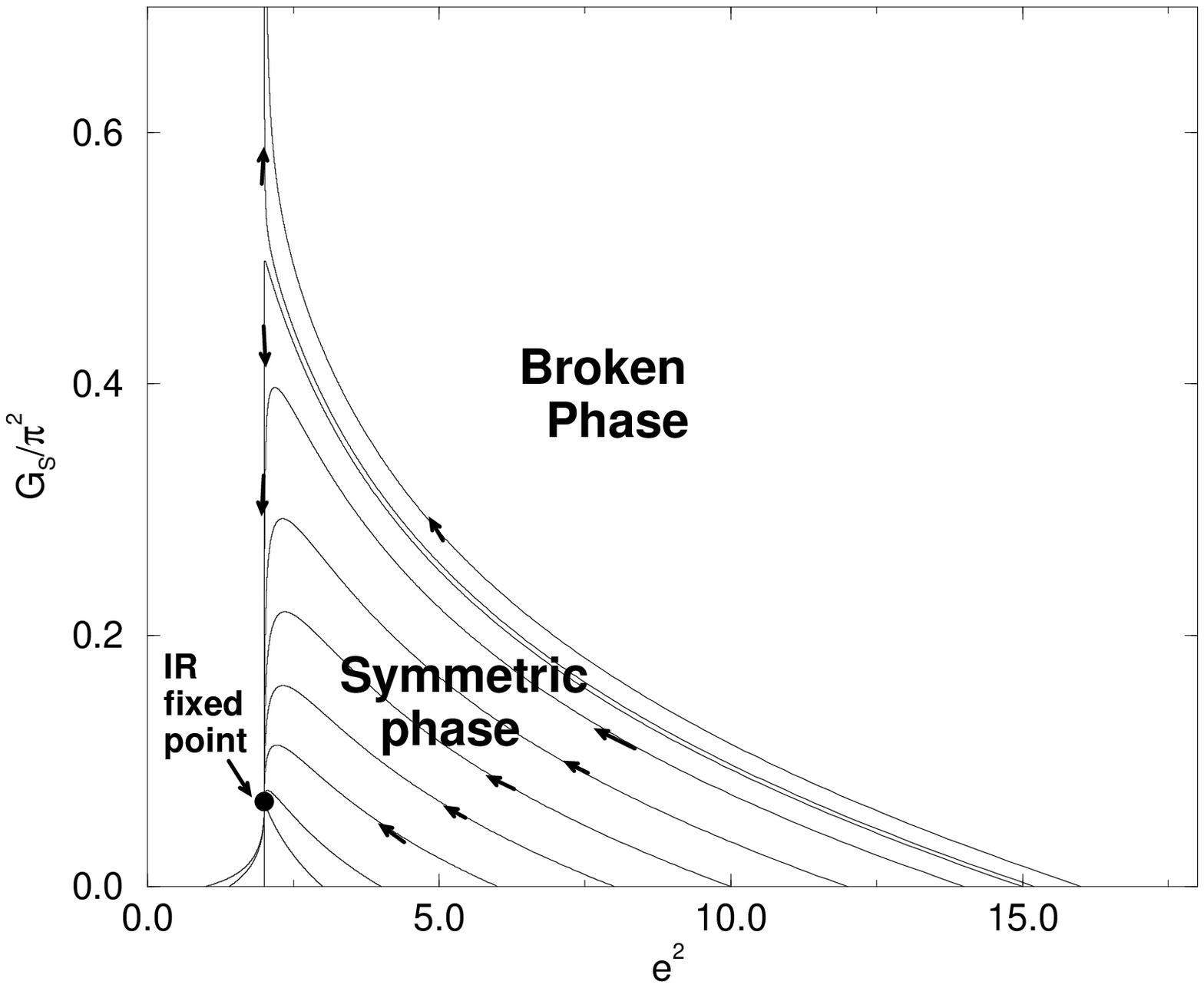}
\caption{RG flows for $\QED$ with $N=4$ flavors in the LPA
are projected on the $(e^2, \Gs )$-plane.}
\end{minipage}
\hspace{5mm}
\leavevmode
\begin{minipage}[t]{60mm}
\epsfxsize=1.0\textwidth
\epsffile{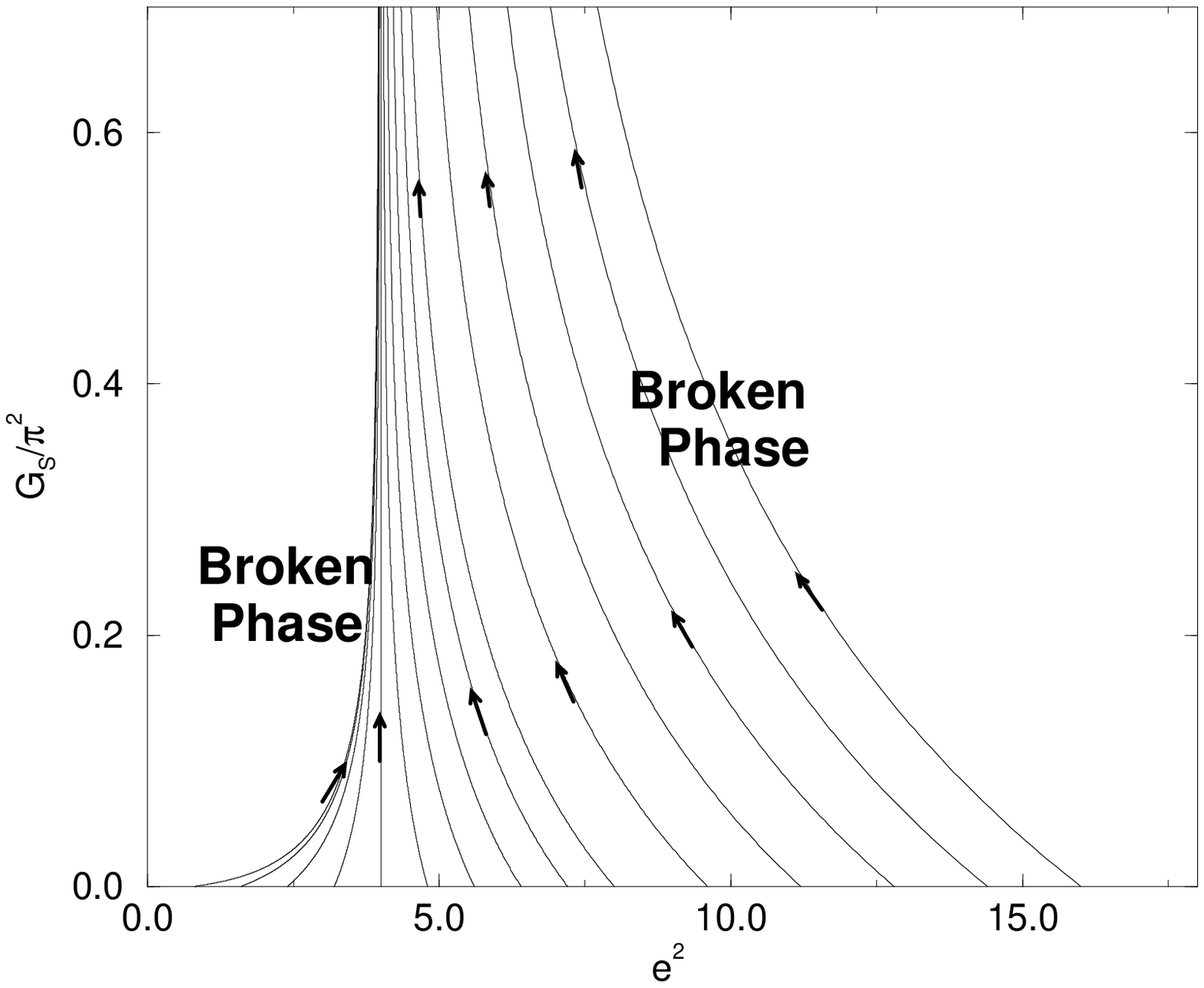}
\caption{RG flows for $\QED$ with $N=2$ flavors in the LPA
are projected on the $(e^2, \Gs )$-plane.}
\end{minipage}
\end{center}
\end{figure}
Comparing Fig.~2 and Fig.~3 with Fig.~4 and Fig.5, we notice that there is no
qualitative difference between them.
This critical number of flavor is also evaluated by the numerical 
calculation and found to be $\Ncr \simeq 3.1$.
This RG analysis naively indicates that $\QED$ becomes a scale invariant
theory, if the flows are absorbed into the IR fixed point.
This is possible only for $N > \Ncr$.

\subsection{Parity broken phase}
In Fig.~6 and Fig.~7, the RG flows on the IR fixed point for the gauge
coupling are shown in $(\Gp,\Gs )$-plane in the case of $N=4$ and $2$
respectively.
It is seen that there are three distinct phases for $N=4$, while
the symmetric phase, where the flows are absorbed into the IR fixed
point, vanishes for $N=2$.
The transition from three phases to two phases also occurs 
at $N=\Ncr$.
The chiral symmetry is supposed to be broken in the upper phase and
the parity symmetry is broken in the right phase.
We cannot conclude the broken symmetries from these RG flows.
However we may find the good evidence by the argument discussed in the
next section.
\begin{figure}[h]
\begin{center}
\begin{minipage}[t]{60mm}
\epsfxsize=1.0\textwidth
\leavevmode
\epsffile{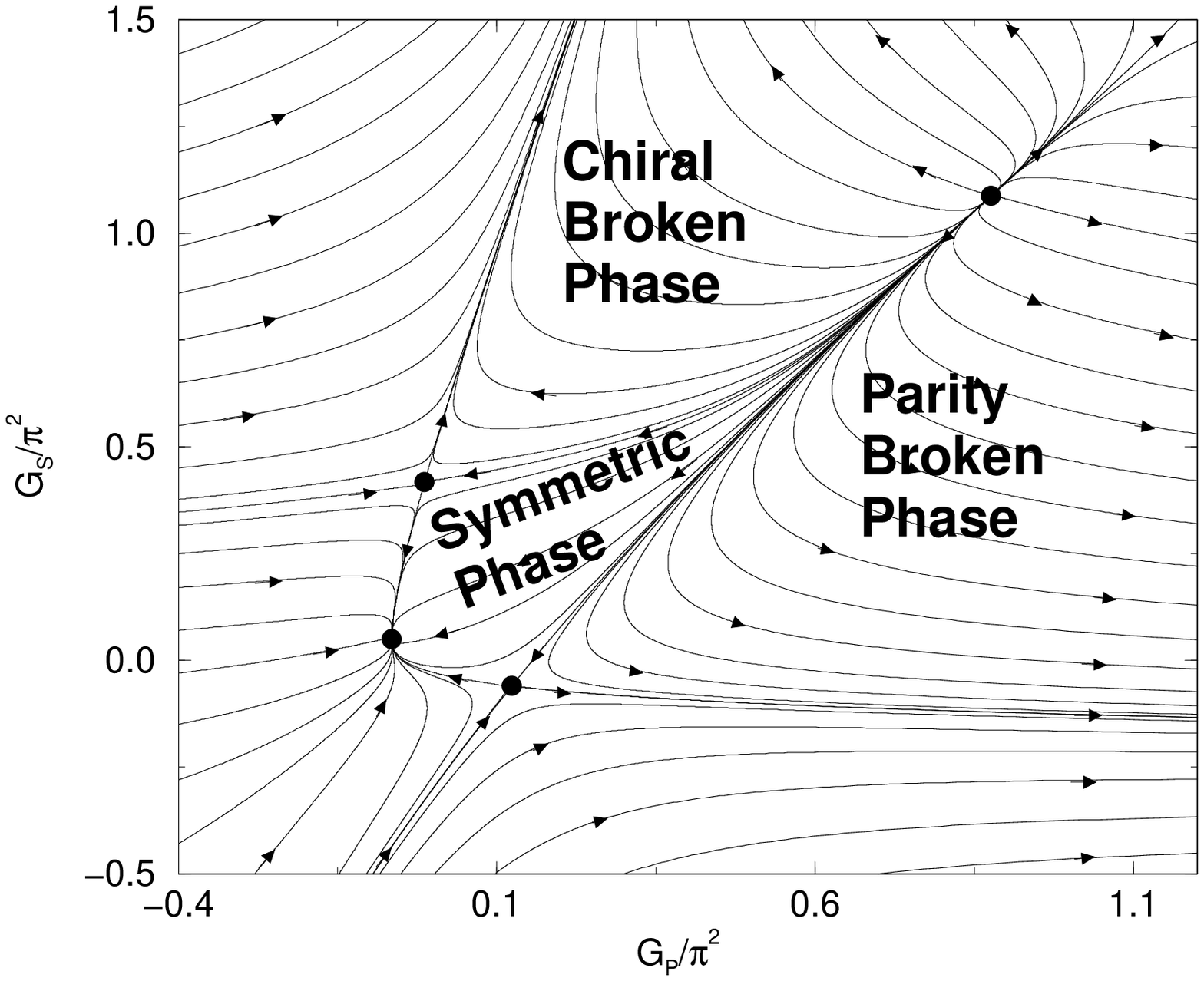}
\caption{RG flows for $\QED$ with $N=4$ flavors in the LPA
are projected on the $(\Gp, \Gs )$-plane.}
\end{minipage}
\hspace{5mm}
\leavevmode
\begin{minipage}[t]{61mm}
\epsfxsize=1.0\textwidth
\epsffile{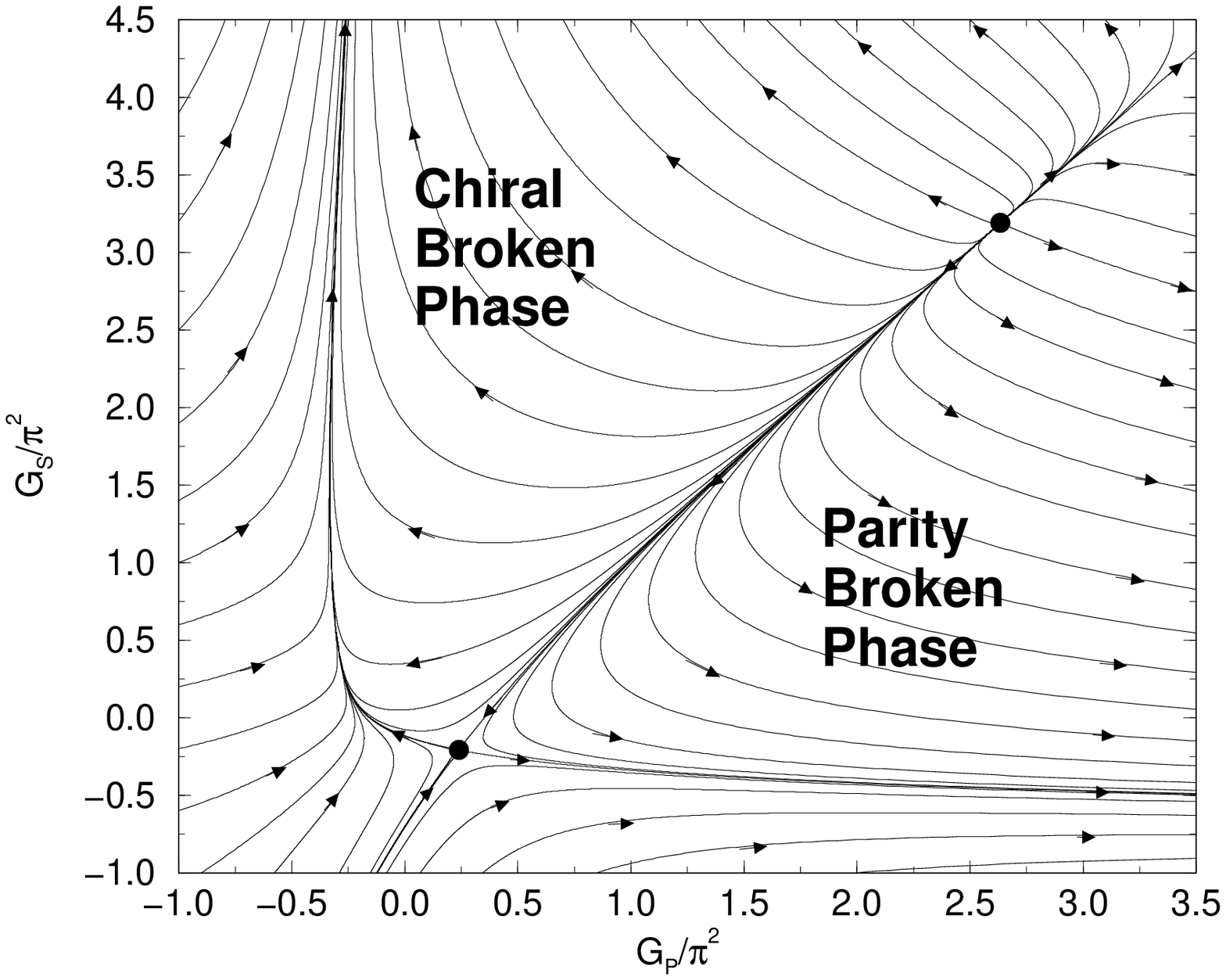}
\caption{RG flows for $\QED$ with $N=2$ flavors in the LPA
are projected on the $(\Gp, \Gs )$-plane.}
\end{minipage}
\end{center}
\end{figure}

Indeed the RG flows for $\QED$ do not enter the parity broken phase.
This corresponds to the Vafa-Witten theorem.
However the phase structure found here shows that the parity symmetry can be
spontaneously broken for the models with the bare four-fermi interactions.
This also coincides with the results obtained by the SD methods \cite{4fermiQED}.
\footnote
{In ref.~\citen{4fermiQED}, however,
the four-fermi interactions added to $\QED$ are not invariant under the  $U(2N)$ symmetry,
which is in contrast to our analysis.
}
Note that we find a tri-critical fixed point at the edge
of the phase boundary between the chiral broken phase and the parity
broken phase.
The tri-criticality indicates that the transition between these
two phases becomes first order beyond this edge.
This point seems to deserve for further study in the NPRG point of
view.
Related with this problem it would be also interesting to
evaluate the anomalous dimensions for
fermion composites, $\bar{\psi}\psi$, $\bar{\psi}\tau\psi$,
$\Os$, $\Op$ and so on at the fixed points.
RG approach is just suitable for such purposes.

At the last of this section 
we would like to stress the advantageous points of the NPRG analyses,
specially in comparison with the SD approaches.
First it is necessary in the SD approach to examine the generalized
models 
in order to explore the whole phase diagram at last.
Whenever we derive the gap equations, we must assume the broken 
symmetries apriori and try some appropriate order parameters.
Also we have to solve the coupled gap equations for them
in the case that there are several non-trivial phases.
For $\QED$ we take care of the parity even mass and the parity odd 
mass.
Therefore the equations become rather complicated to be solved 
even numerically.
Thus the analysis will be rapidly harder in performing 
overall survey of the phase structures and in improving the approximations.
In sharp contrast with such difficult situations, we realize through
the above RG analysis that the NPRG method enables us to explore the whole 
phase diagrams quite easily. Moreover it is not necessary to care about 
the broken symmetries.

\section{Broken symmetries}
So far it has been seen that the RG flows of the four-fermi couplings
clarify the phase boundaries and also the fixed points for $\QED$ and
its generalizations.
In this section we discuss the way to find out the broken symmetry in
each phase.
Our strategy is as follows. The theories belonging to the same phase, or the
same universality class, are supposed to have the common symmetry.
Therefore if we are able to find a much simpler model belonging to the
phase concerned, then we may well examine the dynamical symmetry breaking in 
this model instead of the general theories.
In our case the RG flows show that the three phase structure
remains for the pure four-fermi
theories ($e^2=0$) and also that both of the chirally broken and the parity 
broken phases are connected to those in the whole three dimensional 
theory space.
So we may examine the broken symmetries by using the four-fermi theories.

For the explicit calculation we adopt the so-called auxiliary fields
method.
We shall introduce the following two auxiliary fields in the bare four-fermi
theories:
\begin{equation}
\phi = -\left(\Gp + \frac{\Gs}{2N}\right)(\chibar_I \chi^I), \quad
M^{I}_{J}+\frac{\Gs}{\Gs + 2N\Gp} \delta^{I}_{J}\;\phi = -\Gs (\chibar_J \chi^I).
\end{equation}
Note that $M$ is a traceless hermitian matrix field.
The bare lagrangian is then rewritten as
\begin{equation}
{\cal L}_b = \chibar \left[\Slash{\partial}+ M + \phi\right] \chi + 
\frac{1}{2\Gs} \mbox{tr} \; M^2 + 
\frac{N}{\Gs + 2N \Gp} \phi^2.
\label{auxlagrangian}
\end{equation}
To obtain an effective potential, we calculate only a one-loop
correction for the fermions.
Although this simple approximation is much rougher than the non-ladder 
one adopted in the section 3.4, it would be competent for our present
purpose.
Our effective potential becomes then
\begin{equation}
V_{\mbox{\scriptsize eff}}(M, \phi) = 
\frac{1}{2\Gs} \mbox{tr} \; M^2 + 
\frac{N}{\Gs + 2N \Gp}\phi^2 - 
\int \frac{d^3 p}{(2\pi)^3} \mbox{ln det}\; 
\left[i \Slash{p} + M + \phi \right]
\label{effpotential1}
\end{equation}
All we have to do is to search for the absolute minimum of this effective
potential.

We here investigate only the case of $N=1$ for the sake of simplicity.
The matrix field $M$ may be restricted as 
\begin{equation}
M = \left( \begin{array}{cc} d & 0 \\ 0 & -d \end{array} \right).
\end{equation}
After evaluating the last term in Eq.~$(\ref{effpotential1})$, 
we obtain
\begin{equation}
V_{\mbox{\scriptsize eff}} (d, \phi) = 
\frac{d^2}{\Gs} +
\frac{\phi^2}{\Gs + 2 \Gp} - f(-d+\phi) -f(d+\phi),
\label{effpotential2}
\end{equation}
where
\begin{equation}
f(x) = \frac{1}{6\pi^2} 
\left[\mbox{ln}\; (1+x^2) + 2x^2 -2x^3 \mbox{arctan}\; 
\left(\frac{1}{x}\right)\right].
\end{equation}
$f(x)$ behaves as $ x^2/2\pi^2 $ at $ x \sim 0 $ and ln$ |x|/3\pi^2 $ 
at $ x \sim
\pm \infty $.

Let us consider the broken symmetry from the effective potential.
If the field $d (\phi)$ acquires a vacuum expectation value, 
then the chiral (parity) symmetry is broken down.
$f(-d+\phi)+f(d+\phi) $ reaches its maximal value when $-d+\phi=d+\phi$ or
$-d+\phi=-(d+\phi)$, {\it i.e.} $\phi=0$ or $d=0$.
Therefore the effective potential minimum must be on 
the lines, $\phi=0$ or $d=0$, which means these symmetries are
never broken down simultaneously.
Taking account of the first two terms in Eq.$(\ref{effpotential2})$,
the chiral (parity) symmetry breaking is impossible if $\Gp$ is positive
(negative).
When we make gap equations by differentiating the effective potential
with respect to each field, we find 
if $\Gp > 0$ and $\Gs + 2\Gp > \pi^2$, then the parity symmetry is
broken down, and 
if $\Gp < 0$ and $\Gs > \pi^2$, then the chiral symmetry is
broken down.

\section{Conclusions}
We investigated the phase structure of dynamical symmetry
breaking in $\QED$ with $N$ four-component massless fermions
in the NPRG framework.
The RG flows in the three dimensional theory space spanned 
by ($e^2,\Gs, \Gp$) were explicitly analyzed in the primitive
approximation scheme, however, exceeding the ladder one.
The beta functions for these couplings were gauge parameter independent.
In this RG analysis the theories with the general
four-fermi interactions invariant under the symmetries of $\QED$,
$U(2N)$ and parity, were also automatically incorporated.

>From the RG flows we found that the theory space was divided
into three phases for $N > \Ncr$, while the symmetric phase
disappeared for $N < \Ncr$. The critical flavor number was
numerically estimated as $\Ncr \simeq 3.1$.
The phases were classified into the chiral symmetry broken,
the parity broken and the scale invariant one.
We elucidated the broken symmetries in each phase by
examining the four-fermi theory belonging to the phase.
Also the tri-critical fixed point was found to exist at the
edge between the chirally broken phase and the parity broken
one.

It was also seen within this approximation that the parity symmetry never
be spontaneously broken in $\QED$ with four-component fermions. 
This result was consistent with the Vafa-Witten theorem.
For $N < \Ncr$ the chiral symmetry was broken irrespective of the
gauge coupling, while the critical gauge coupling dividing
into chirally  broken and unbroken phases was found
to exist for $N > \Ncr$.
Since only the asymptotically free theories are examined in these
analyses, these were also consistent with the known results obtained by
solving the SD equations and by the lattice simulations 

Through these analyses it was demonstrated that the NPRG 
offered us powerful and simple methods to explore the phase
structures.
The approximation scheme could be systematically improved
in principle.
However it has been the hard problem to maintain the
gauge invariance non-perturbatively in the ERG method.
Indeed our treatment for the gauge beta function was quite
poor.
For $N$ not large the results were not totally confidential
because the fixed points appears at the strongly coupled region.
It should be hasten to develop the ERG framework for the
gauge theories.

As final remarks we would like to mention other possible applications
of such a NPRG approach.
The similar dynamical phenomena to $\QED$ has been known to occur also
in four dimensions.
If we consider QCD with appropriately many flavors, then the IR fixed point
is found to exist \cite{Banks}.
There must be the critical number of flavors 
where the chiral symmetry breaking
start to occur \cite{conformal}.
The essential mechanism of this transition may be
understood by disappearance of the fixed point.
Another example may be found in the high density QCD.
Recently the study of QCD with finite density has been revived 
and it is also believed that the so-called color superconducting phase
exists \cite{denseQCD}.
Actually most studies to these phenomena have been done by considering
the SD equations.
Most probably the NPRG will be useful also in these area.

%%%%%%%%%%%%%%%%%%%%%%%%%%%%%

\section*{Acknowledgements}

The authors would like to thank K.~I.~Kondo, H.~So, M.~Tomoyose and specially 
Porf. V.~A.~Miransky for valuable discussions and comments.

\end{document}